\documentclass[manuscript]{acmart}
\usepackage{subcaption}
\usepackage{booktabs}   
\usepackage{caption}    
\usepackage{tabularx}   
\usepackage{multirow}   
\usepackage[normalem]{ulem}
\usepackage{xurl}
\AtBeginDocument{%
  }

\setcopyright{acmlicensed}
\copyrightyear{2018}
\acmYear{2018}
\acmDOI{XXXXXXX.XXXXXXX}
\acmConference[Conference acronym 'XX]{Make sure to enter the correct
  conference title from your rights confirmation email}{June 03--05,
  2018}{Woodstock, NY}
\acmISBN{978-1-4503-XXXX-X/2018/06}

\begin{document}

\title{FenceXR: AR Movement Replay for Error-Detection Training and Spatially Grounded Feedback}

\author{Avinash Ajit Nargund}
\email{anargund@ucsb.edu}
\affiliation{%
  \department{Electrical and Computer Engineering}
  \institution{University of California, Santa Barbara}
  \city{Santa Barbara}
  \state{CA}
  \country{USA}
}

\author{Amelia Haruka Harrison}
\affiliation{%
  \department{Psychological and Brain Sciences}
  \institution{University of California, Santa Barbara}
  \city{Santa Barbara}
  \state{CA}
  \country{USA}
}

\author{Timothy Robinson}
\affiliation{%
  \institution{University of California, Santa Barbara}
  \city{Santa Barbara}
  \state{CA}
  \country{USA}
}

\author{Tobias Höllerer}
\affiliation{%
  \department{Computer Science}
  \institution{University of California, Santa Barbara}
  \city{Santa Barbara}
  \state{CA}
  \country{USA}
}

\author{Misha Sra}
\affiliation{%
  \department{Computer Science}
  \institution{University of California, Santa Barbara}
  \city{Santa Barbara}
  \state{CA}
  \country{USA}
}
\renewcommand{\shortauthors}{Trovato et al.}
\newcommand{\sys}{FenceXR}
\begin{abstract}
Recognizing technical errors in movement is a perceptual skill important to motor learning, but it is challenging for beginners in fast, complex sports like fencing to develop it. A coach's attention is scarce, live demonstrations vary from repetition to repetition, and video review is limited to whatever camera angle was used to record it. Coaches and advanced fencers reviewing a recording face a related problem. They can see an error, but have no way to anchor their feedback to the movement itself, and are left describing it in words the learner must map back onto their own body. We present FenceXR, an augmented reality system that reconstructs 3D movement replays from monocular smartphone video to address both problems. A \textit{Trainee} module trains novices to detect common lunge errors while a \textit{Reviewer} module lets coaches and advanced fencers attach text or voice annotations to a specific joint and moment in a replay, which can be shared asynchronously with a trainee. In a study with 18 novice fencers, unaided error-detection accuracy rose from near-chance (37.5\%) before training to 64.1\% after a single session, with interviews showing a shift from broad visual scanning toward targeted inspection of specific joints and their timing. In a video-based study with four fencing experts, all four viewed the \textit{Reviewer} module as a valuable complement to their existing coaching tools, particularly for feedback that is difficult to convey through standard video. We end with a discussion of implications for designing AR systems that ground movement-based training and feedback in the movement itself.
\end{abstract}

\ccsdesc[500]{Human-centered computing~Interactive systems and tools}
\ccsdesc[500]{Human-centered computing~Mixed / augmented reality}

\keywords{Motor Training, Error-detection, Sports, AR, Fencing, Observational Learning}
\begin{teaserfigure}
  \includegraphics[width=\textwidth]{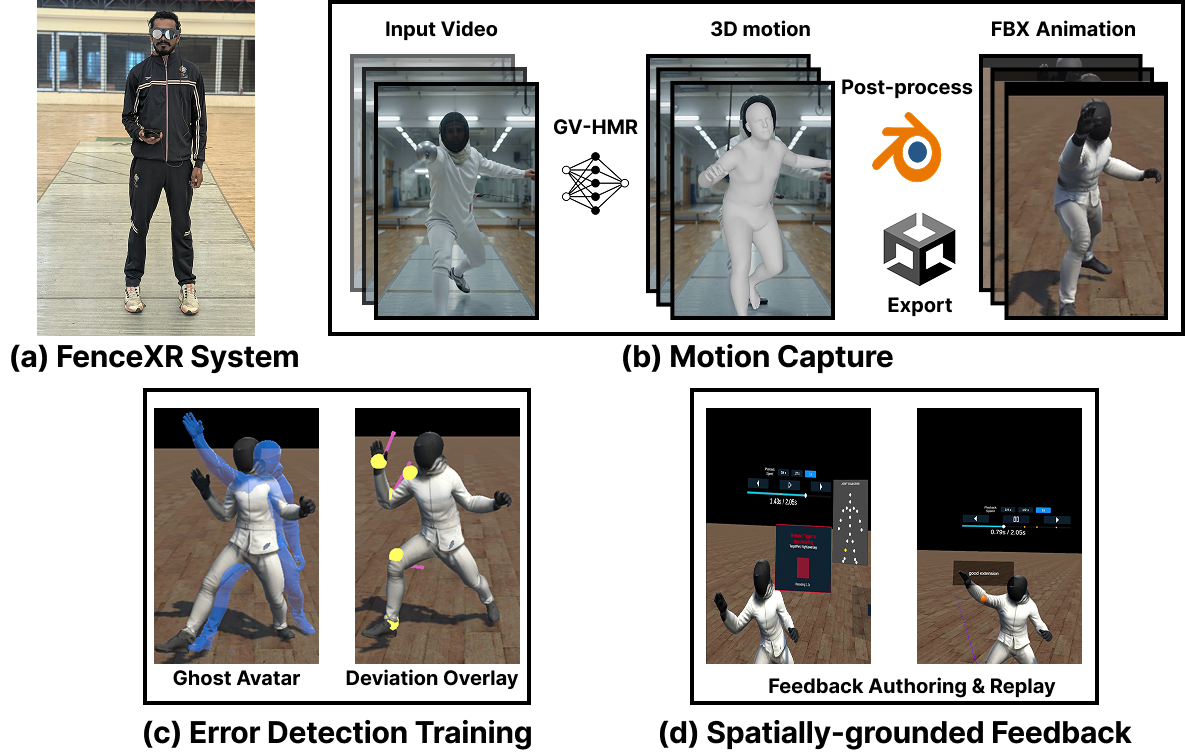}
  \caption{\sys~is an AR movement replay system that enables error-detection training and spatially grounded feedback for fencing. (a) A user wearing a Magic Leap 2 running \sys, which is built on (b) a monocular video-to-3D reconstruction pipeline: smartphone video is converted to a mesh using GV-HMR~\cite{shen2024world}, post-processed via the Blender Python API, and exported to Unity as an FBX, building a reusable library of fencing lunge movements. This library powers \sys's two modules: (c) the \textit{Trainee} module trains error-detection skill using one of two visualization modalities, \textit{Ghost Avatar} or \textit{Deviation Overlay}, and (d) the \textit{Reviewer} module that enables coaches and experts to provide spatially grounded feedback to their students.}
  \Description{}
  \label{fig:teaser}
\end{teaserfigure}

\received{20 February 2007}
\received[revised]{12 March 2009}
\received[accepted]{5 June 2009}

\maketitle

\section{Introduction}

The ability to recognize technical errors in one's own and other people's movements is central to motor skill learning~\cite{seidler2013}. This perceptual skill supports self-correction, since learners who detect and correct their own errors achieve better performance outcomes~\cite{duke2009s}, and it underpins effective peer feedback, which benefits motor skill acquisition~\cite{trabelsi2025direct}. Error-detection is a trainable skill and structured observation of others' movement can develop error-detection ability as effectively as physical practice of the movement itself~\cite{blandin2000}. However, developing it requires the opportunity to see an error clearly, and receive guidance about which movement features to attend to. 

Beginners in fast, biomechanically complex sports such as fencing~\cite{yiou2001complex} often have neither. Errors in fencing movements are difficult to perceive as they occur, and beginners often lack the perceptual expertise to know which features to attend to~\cite{sigrist2011, abernethy1987}. Coaches address this through live demonstration of correct and incorrect execution, but a coach's attention is a scarce resource shared across a class, and most of a beginner's practice, drilling and sparring with peers, may happen without direct coach input. Opportunities to observe technique are further limited in smaller clubs, where learners repeatedly train with the same small group of peers. Even when a coach does demonstrate, natural variation between repetitions can make it difficult to isolate the defining features of an error from any single demonstration~\cite{preatoni2013movement}. Video replay lets learners revisit a movement outside the training session, but videos do not indicate what to look for. They are typically limited to one or a few fixed camera angles. The vantage point needed to see a given error, a lateral view for lower-body timing, a frontal view for shoulder alignment, may not be among them.

Feedback from a coach adds knowledge of performance and information about the movement pattern itself, which self-detection alone cannot supply. But because coach attention is concentrated in the training session, movement performed outside it can typically be reviewed only asynchronously, from a recording. Coaches and advanced fencers reviewing such a recording can already see the errors but they lack is a means of attaching what they see to the movement. Without spatial annotation tools they must describe errors verbally or in writing, leaving the learner to map those descriptions back onto their own body~\cite{wen2024augmented}. Both problems stem from the same underlying issue that information about movement is decoupled from the movement it describes. What differs is what each user requires to close that gap. Beginners have not yet formed a perceptual model of correct and erroneous execution, and require their attention directed to the features that distinguish one from the other while reviewers already possess this mental model and require a means of externalizing it.


Extended reality (XR) offers a way for closing this gap by rendering movement as an interactive 3D replay that can be inspected from any viewpoint at one-to-one scale, replayed identically, and paused on-demand at important moments. Advances in monocular video-to-3D human reconstruction~\cite{goel2023humans, shen2024world, li2021hybrik, li2025hybrik} allow such replays to be produced from a single smartphone video, removing the need for motion-capture hardware or multi-camera rigs, broadening access to small fencing clubs. Free viewpoint alone, however, addresses access but not attention. A novice given full control over a 3D replay still does not know where to look, and the additional degrees of freedom may even compound the problem. Prior XR systems for motor learning largely target reproducing a movement correctly rather than perceiving errors in it~\cite{peng2025, trainme2024}, while trainer-facing systems that support spatial annotation, most closely Augmented Coach~\cite{wen2024augmented}, depend on multi-sensor volumetric capture. The questions of how to train the perceptual skill of error detection, and how to ground reviewer feedback in the movement itself from commodity capture, remain open.

In this work, we present \sys, an AR system (Figure~\ref{fig:teaser}a) that transforms monocular smartphone video into interactive 3D movement replays supporting both error-detection training and spatially grounded asynchronous feedback for the fencing lunge. \sys~was developed with guidance from our second author, a high-level competitive fencer with experience coaching beginners, who identified the target movement, the error set, and the coaching requirements. A fundamental technique in fencing is the lunge, taught on a beginner's first day and used through elite competition. Within it, we target two common errors in beginners: \textit{shoulder rotation}, where a fencer rolls the shoulder during arm extension and shortens their attacking reach, and \textit{leg-leading}, where the front leg initiates the lunge before the arm extends, giving an opponent time to respond. Shoulder rotation is a postural error visible in a single frame, whereas leg-leading is a timing error that can be assessed only across the full movement. Together they test whether our approach supports detection of both.

Our reconstruction pipeline converts smartphone video into segmented 3D lunge replays using off-the-shelf monocular-to-3D models with custom post-processing (Figure~\ref{fig:teaser}b). Built on this pipeline, two modules provide the scaffolding each user requires. Consistent with evidence that observing others' movement can build error-detection skill as effectively as physical practice~\cite{blandin2000}, the \textit{Trainee} module (Figure~\ref{fig:teaser}c) trains this skill on canonical exemplar replays of correct and erroneous lunges performed by an expert instead of the trainee's own movement. It provides two visualizations, \textit{Ghost Avatar} and \textit{Deviation Overlay}, that expose differences from a reference lunge in different ways, together with separate training and testing modes that distinguish scaffolded practice from unaided assessment. The \textit{Reviewer} module (Figure~\ref{fig:teaser}d) lets coaches and advanced fencers inspect a reconstructed movement in AR and attach text or voice annotations to a specific joint at a specific moment, which are persisted separately from the movement and can be replayed asynchronously by a trainee.

We evaluated the \textit{Trainee} module with 18 novice fencers, who classified unaugmented replays of an expert performing correct and erroneous lunges, before and after a single training session. Before training, participants recruited from fencing clubs performed close to chance on a three-alternative classification task ($M=37.5\%$; chance $=33.3\%$), confirming that these errors are difficult to perceive without guidance. After training, unaided accuracy rose to $64.1\%$ ($p<.001$, $\eta^2_p=.52$). Participants also left fewer trials unanswered and reported higher decision confidence. Both visualization modalities trained with an augmented, labeled replay, so we attribute this gain to training with \sys as a whole rather than to either visualization alone. The accuracy gain was significant in the \textit{Deviation Overlay} modality but not in \textit{Ghost Avatar}. The difference between modalities, however, was not itself significant at this sample size, so we treat the visualization comparison as exploratory. Interviews show how the gain arose as participants adopted viewpoint strategies to match to each error, used the two visualizations in distinct ways, and shifted from holistic scanning toward structured inspection of specific joints and inter-joint timing. We evaluated the \textit{Reviewer} module through a video-walkthrough study with four fencing experts, who rated the system's slow-motion replay and joint-anchored annotation capabilities favorably and considered it a valuable complement to their existing coaching tools, particularly for surfacing errors not visible in standard video and delivering feedback students could more easily understand.

This paper makes the following contributions:

\begin{itemize}
  \item A monocular smartphone-video-to-3D reconstruction pipeline for fencing, validated by confirming that the kinematic markers separating correct execution from shoulder rotation and leg-leading errors survive reconstruction.
  \item \sys, an AR system that grounds movement information in the movement itself, providing attention-directing visualizations for novices and joint- and time-anchored annotation authoring for reviewers.
  \item Empirical evidence that the AR training session improved beginner fencers' unaided detection of both a postural and a timing error, together with a qualitative account of their learning strategies.
  \item Preliminary evidence from a video-walkthrough study with fencing experts that joint-anchored, spatially grounded annotations are seen as a valuable complement to existing remote-coaching tools.
\end{itemize}

\section{Related Work}
Our work is related to the use of XR, specifically the analysis and annotation of 3D movement replays, in the context of observational learning for motor skill acquisition. Several works have investigated XR-based feedback for applications such as dance~\cite{kyan2015a, trajkova2018}, sports~\cite{ma2024, ikeda2019, wu2025, lenaour2019a}, or exercises~\cite{tang2015, escalona2020eva}, but these systems largely support performing a movement correctly rather than helping learners perceive and diagnose errors in technique, whether their own or others'. Similarly, while a few systems have explored trainer-facing spatial annotation in XR for sports such as basketball~\cite{wen2024augmented} or badminton~\cite{lin2024}, no prior work has examined how such annotation could be adapted for fencing, where coaches rely on spatially grounded feedback in person to help learners detect errors. In this section, we provide a brief overview of prior work to contextualize the contributions of our research.

\subsection{Error-Detection in Motor Learning}
Motor learning requires progressing through a cognitive stage, in which learners acquire and remember the overall pattern of a movement, and an associative stage, in which learners refine that movement by detecting and correcting their own errors~\cite{fitts1967human}. Feedback from a coach or coaching system is critical throughout both stages, but the associative stage additionally depends on the learner's ability to notice when and how their movement deviates from correct technique. Prior work has shown that watching someone else perform a motor skill can help a learner develop error-detection and correction skills as effectively as physically practicing the skill themselves~\cite{blandin2000}, suggesting that structured observation can meaningfully develop error-detection skill independent of a learner's own physical repetition. 
Despite this, majority of existing XR systems for motor learning have primarily focused on helping users train and refine the movement~\cite{yu2024, diller2024}, rather than on developing the perceptual skill of detecting errors in that movement. Our system instead treats error-detection as a perceptual skill that can be trained directly, using structured observation of others' movement to build this skill.

\subsection{XR Systems for Motor Skills}

XR systems for motor skill training have converged on a few recurring feedback mechanisms. Most rely on real-time comparison between the learner's movement and a reference or expert performance, visualized either as a superimposed 3D overlay~\cite{ma2024, lenaour2019a, kyan2015a} or through an augmented mirror providing visual and textual correction~\cite{tang2015,trajkova2018}. Some systems use post-movement replay and visualization rather than concurrent overlay, giving learners a similarity score or reviewable capture of their attempt against a reference performance ~\cite{oshita2018, wu2025, watanabe2026}. Across all these approaches, however, the underlying goal is the same: helping the learner reproduce a correct movement more accurately, rather than training the learner to independently recognize when and how a movement deviates from correct technique.

A smaller body of work has explored designing XR systems grounded specifically in observational learning principles. Karate novices using a smartphone-based VR application to observe correct technique showed improved movement execution, supporting observational learning as a viable mechanism for XR-based training~\cite{ohl2019improvement}. Closer to our work, FencBuddy~\cite{peng2025} helps fencers train their offensive actions and distance perception using VR, scoring the similarity between their actions and an expert's via skeleton tracking, and offers unannotated replay for self-review. Train Me~\cite{trainme2024} explores mobile AR capture and replay for self-training across sports, aiming to reduce reliance on an on-site coach. Both systems, however, train learners toward reproducing correct execution rather than toward independently detecting and diagnosing errors in a movement, and neither incorporates a coach-facing feedback mechanism. We instead treat error-detection as a perceptual skill that can be trained directly, and design our system to help learners spot errors accurately. 

Most XR systems for trainers focus on helping coaches communicate tactical ideas by exploiting the spatial dimension~\cite{lin2024, liqi2024, chu2022}. A smaller number of systems enable coaches to provide spatially grounded guidance and feedback to learners on their technique~\cite{thoravikumaravel2019, mrcoach2025}. Among these, Augmented Coach is closest to our work~\cite{wen2024augmented}. It is a trainer-oriented XR system that volumetrically reconstructs an athlete's movement from multiple Kinect sensors, letting coaches review the reconstruction and attach joint-wise text or voice annotations. Our \textit{Reviewer} module supports the same core interaction, but reconstructs movement from a single monocular smartphone video rather than a multi-sensor volumetric capture rig, substantially lowering the setup and hardware required to capture and review a movement in XR.
  


\section{System Design and Implementation}
 ~\sys is an AR system that trains novices to better detect errors in their lunge and coaches to provide spatially grounded feedback to their students. It consists of three components: a motion capture module that reconstructs 3D movement from monocular smartphone video, a \textit{Trainee} module that trains novices to detect technique errors through augmented movement replay, and a \textit{Reviewer} module that lets coaches and experts annotate and share feedback on the same reconstructed movements.

\subsection{Design Rationale}~\label{sys:rationale}
The design of \sys~was guided by our second author, a fencing expert with coaching experience. Their insights from coaching beginners directly shaped which movement and errors the system targets and how both the \textit{Trainee} and \textit{Reviewer} modules were designed.

\paragraph{\textbf{Why the lunge?}}
Our co-author identified the lunge as fencing's most fundamental attacking technique. It is taught on a beginner's first day alongside basic footwork, and is a primary technique used all the way through elite-level competition, similar to learning a jab or cross in boxing. Its early, central role in training means that correctly executing the lunge, and identifying errors in it, has a significant impact on a beginner's progress, prompting us to make it the focus for our system.

\paragraph{\textbf{Why shoulder rotation and leg-leading errors?}}
Our co-author identified shoulder rotation and leg-leading as two of the most common beginner errors. Shoulder rotation results from a beginner lacking the strength to hold their weapon steady through a full arm extension. They compensate by tightening and rolling the shoulder, which shortens their attacking reach--a disadvantage in fencing, where landing a hit from maximum distance is advantageous. Leg-leading errors, where the front leg initiates the lunge before the arm extends, could stem from everyday reaching habits, where people often reduce body-target distance before extending their arm. This tendency is compounded in beginners, who tend to focus on their footwork before they can coordinate it with arm extension. Leading with the leg introduces a timing vulnerability, giving an opponent more time to defend or make a counterattack. 

These two errors differ in how much information is needed to detect them. The shoulder rotation is visible from a single still frame, whereas leg-leading can only be assessed by observing the full movement. Together, they illustrate two distinct error types relevant to lunge technique: \textit{postural errors}, incorrect body pose at a single moment, and \textit{timing errors}, incorrect sequencing across the movement. We use these two errors as examples to examine whether our visualizations support error-detection training across both types.

\paragraph{\textbf{Why novices struggle to detect these errors?}}
Our co-author explained that novices often miss these errors not because the errors are invisible, but because novices lack an internal template of correct and incorrect technique, and are unlikely to notice a deviation without being told what to look for. Building this template typically requires substantial experience, both in performing the technique and in observing others perform it, exposure that is especially limited in smaller clubs where students repeatedly train with the same small group of peers. This motivates a system that supplements novices' limited first-hand exposure with structured, repeatable practice at detecting commonly-made errors.

\subsection{Motion Capture}~\label{sys:mocap}


We used a monocular video-to-3D reconstruction pipeline as the primary means of transferring recorded fencing movements into AR. We evaluated several off-the-shelf monocular human motion reconstruction models, including 4D-Humans~\cite{goel2023humans}, VIBE~\cite{kocabas2020vibe}, GVHMR~\cite{shen2024world}, and HybrIK-X~\cite{li2025hybrik}. For each model, we used the predicted SMPL parameters (shape coefficients, root translation, global orientation, and joint rotations) to compute the corresponding mesh via forward kinematics using the SMPL~\cite{smpl2015} PyTorch layer. Predicted sequences were converted from SMPL's $Y$-up to Blender's $Z$-up coordinate system for import.

The resulting mesh was translated so that the feet made proper ground contact in the initial frames, adjusting pelvis translation each frame based on the minimum vertex height to correct foot-sliding and floating artifacts common to monocular reconstruction, especially when the motion is towards the camera~\cite{wu2025}. The post-processed output was imported into Blender using its Python API\footnote{\url{https://docs.blender.org/api/4.3/index.html}} and written directly to Blender's internal F-Curves in a single batch operation, rather than through per-frame keyframing calls, to reduce export time. Each sequence was automatically segmented into individual lunge repetitions using a smoothed lateral neck-joint signal to detect repetition boundaries, with each repetition stored as a separate animation on the reconstructed mesh and exported as an FBX file for import into our AR application.

The authors and pilot testers assessed the resulting sequences for temporal smoothness, foot-ground contact, and displacement accuracy across more than 15 minutes of recorded lunge footage. Based on this assessment, we selected GVHMR~\cite{shen2024world} as our reconstruction model. 
 
\subsection{Trainee Module}~\label{sys:trainee}
The \textit{Trainee} module trains novices to detect technique errors in a lunge by comparing it against a fixed expert reference lunge. Trainees view movement replays through a shared interface that lets them freely rotate the replay avatar and pause or resume playback to inspect poses in detail. To support this training, we implemented two visualization modes, \textit{Ghost Avatar} and \textit{Deviation Overlay} (\ref{sys:trainee_viz}), that highlight differences between the exemplar movement and the reference in different ways, and two interface modes, \textbf{training} and \textbf{testing} (\ref{sys:trainee_inter}), that control whether this visualization and the ground-truth error label are shown or hidden, allowing trainees to both learn and assess their own error-detection ability.

\subsubsection{Visualizations}~\label{sys:trainee_viz}
\begin{figure}
    \centering
    \begin{subfigure}[b]{0.49\linewidth}
        \centering
        \includegraphics[width=\textwidth]{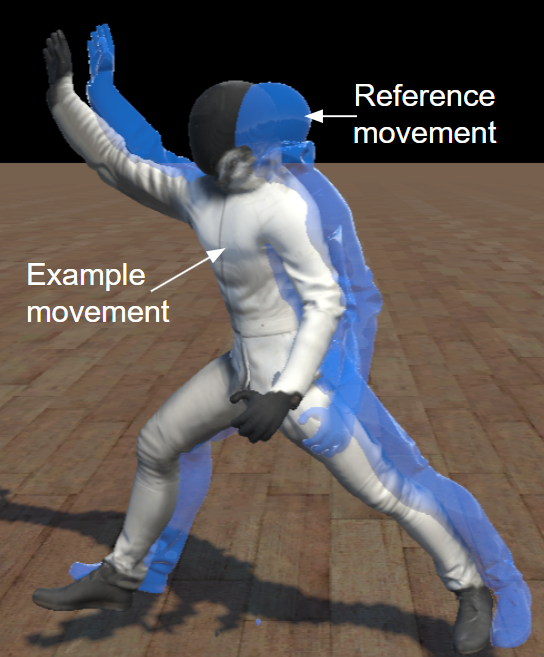}
        \caption{Ghost Avatar}
        \label{fig:ghost}
    \end{subfigure}
    \begin{subfigure}[b]{0.49\linewidth}
        \centering
        \includegraphics[width=\textwidth]{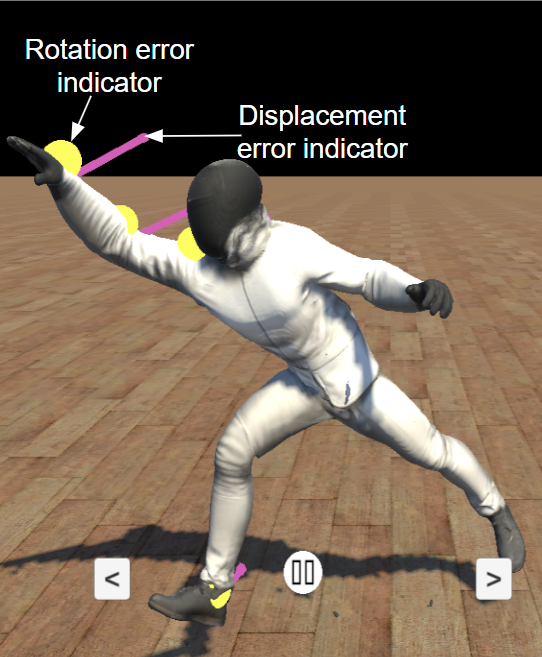}
        \caption{Deviation Overlay}
        \label{fig:traj}
    \end{subfigure}
    \caption{The visualization modalities that are used for error-detection training in ~\sys. In the \textit{Ghost Avatar}, a semi-transparent avatar depicting the reference movement is superimposed on the avatar depicting the exemplar movement. In the \textit{Deviation Overlay} mode annotations indicating joint rotation and displacement errors are attached to the avatar depicting the exemplar movement.}
    \label{fig:viz}
\end{figure}

We implemented two visualizations (Figure~\ref{fig:viz}) to support error detection learning. Before either visualization is generated, the exemplar's movement is temporally aligned to the reference movement using dynamic time warping (DTW) with the joint displacement metric computed over seven joints: the hips, chest, right shoulder, both hands, and both feet. These joints were selected as they are most relevant to the lunge movement. The colors used in both visualizations were selected from a colorblind-accessible palette~\footnote{\url{https://davidmathlogic.com/colorblind/\#\%23FEFE62-\%23D35FB7}}. 

\paragraph{Ghost Avatar} 
This visualization renders the reference movement as a semi-transparent avatar, spatially aligned and superimposed on the DTW-aligned exemplar movement. The design of the Ghost Avatar is inspired by the transparent hand visualizations used for movement guidance in
EGuide~\cite{durr2020} and work by~\citet{lilija2021}.
\paragraph{Deviation Overlay}  This visualization mode overlays deviation indicators directly onto the avatar depicting the exemplar movement. Joints whose rotation differs from the reference beyond a threshold are highlighted with a yellow sphere. Similarly, when displacement between the joints of the exemplar's movement and the reference movement exceeds a threshold, a purple line segment starting from the current joint position to its correct position is used to indicate the discrepancy. Both the rotation and displacement threshold are tuned through pilot testing. 

\subsubsection{Interface}~\label{sys:trainee_inter}     

\begin{figure}
    \centering
    \begin{subfigure}[b]{0.595\linewidth}
        \centering
        \includegraphics[width=\textwidth]{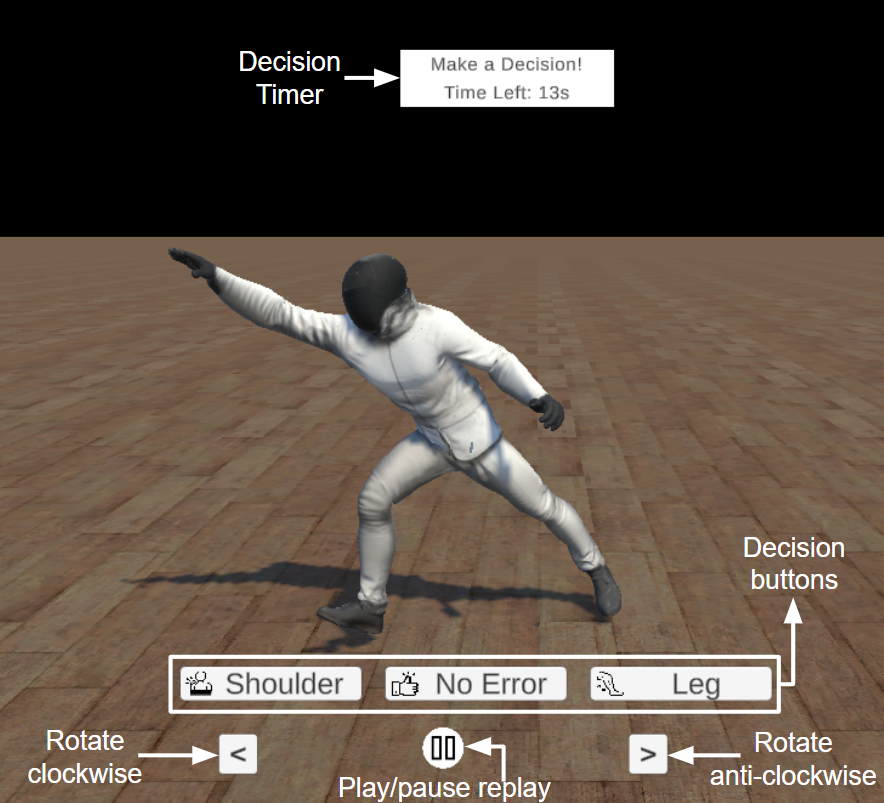}
        \caption{Error-detection Pre-/Post-Test Interface}
        \label{fig:err_test}
    \end{subfigure}
    \begin{subfigure}[b]{0.4\linewidth}
        \centering
        \includegraphics[width=\textwidth]{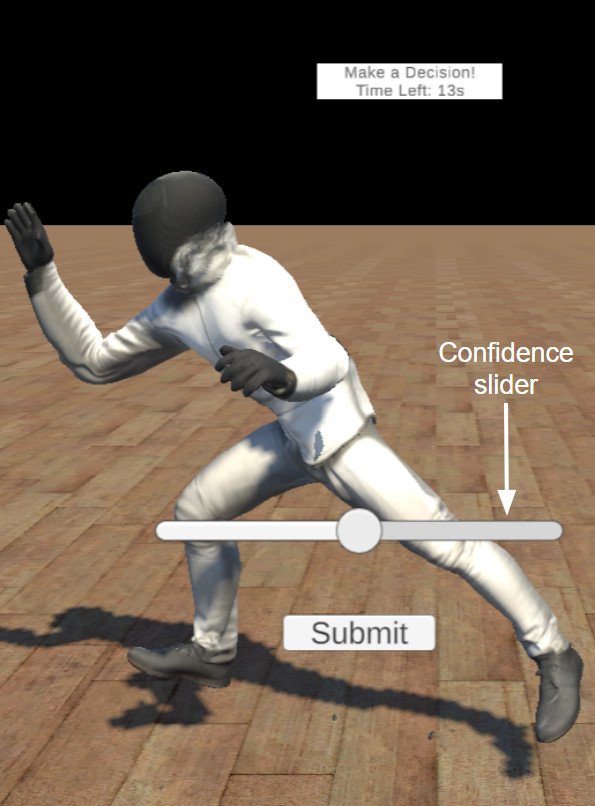}
        \caption{Confidence Slider}
        \label{fig:conf_slider}
    \end{subfigure}
    \caption{(a) The AR interface used for the pre- and post-tests. Participants could rotate the avatar clockwise or anti-clockwise and play or pause the replay to carefully examine movement poses. After classifying the movement as correct or as exhibiting a shoulder rotation or leg-leading error, participants recorded their decision using the decision buttons. A timer above the avatar's head helped them track elapsed time. (b) After making their decision, participants indicated their confidence using a slider (left/1 = not confident, right/5 = confident).}
    \label{fig:test_interface}
\end{figure}

\begin{figure}
    \centering
    \includegraphics[scale=0.4]{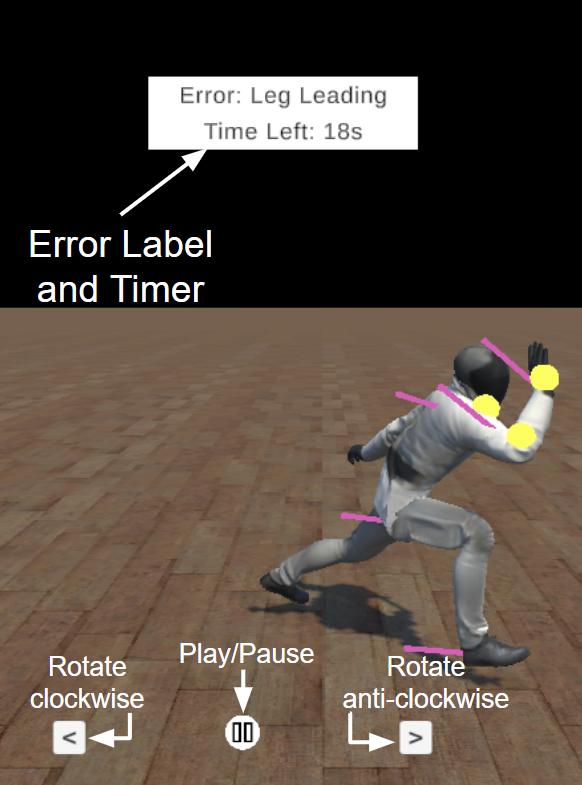}
    \caption{The AR interface used for error detection training. Participants could rotate the avatar clockwise or anti-clockwise and play or pause the replay to carefully examine movement poses. An error label conveys the type of movement being shown. A timer above the avatar's head helped them track elapsed time.}
    \label{fig:training_int}
\end{figure}

During \textbf{training}, each replay is shown with its ground-truth error label (no error, shoulder rotation, or leg-leading) along with the chosen visualization (\textit{Ghost Avatar} or \textit{Deviation Overlay}), allowing learners to see the type of error and where it appears in the movement (Figure~\ref{fig:training_int}). In \textbf{testing} mode, both the error label and visualization are hidden. Trainees instead view an unannotated replay and classify it themselves, selecting one of the three error categories (no error, shoulder rotation, leg-leading) using the on-screen buttons(Figure~\ref{fig:test_interface}a). After making a decision, learners can additionally indicate how confident on a scale $1-5$ they are in their decision using a slider(Figure~\ref{fig:test_interface}b).

\subsection{Reviewer Module}~\label{sys:Reviewer}
The \textit{Reviewer} module lets reviewers author feedback on a fencer's lunge that stays anchored to the movement itself, rather than described separately in writing or speech. It allows reviewers to analyze reconstructed 3D replays of movement and attach text or voice annotations directly to specific joints and moments, addressing the spatial and temporal disconnect between verbal feedback and the recorded performance described in Section~\ref{sys:rationale}. We first describe the module's interface, then the workflow reviews follow to author and view annotations.

\subsubsection{Interface}
Feedback-authoring tools in the \textit{Reviewer} module are anchored spatially around the avatar rather than fixed to the user's head-locked view. The playback control panel is anchored above the avatar and includes a scrubbable timeline, a play/pause button, buttons to rotate the avatar in fixed increments to view it from different angles, and buttons to switch between three discrete playback speeds (0.25$\times$, 0.5$\times$, and 1$\times$ real-time) to support close inspection of fast movement segments. The annotations control panel lets coaches select between pre-loaded movement clips, switch between authoring new feedback and reviewing saved sessions, toggle between text and voice annotation modes, and leave repetition-level feedback via a \textit{General Review} button, in addition to feedback anchored to a specific joint. Reviewers can select a specific joint either by pointing and clicking at the corresponding joint on the avatar or on the mirrored 2D skeletal diagram positioned beside the avatar, which offers a reliable alternative when a joint is occluded in the 3D scene. All spatial panels remain billboarded toward the user's viewpoint as they move around the avatar, keeping controls legible without breaking spatial anchoring to the avatar itself.

\subsubsection{Annotation Workflow}
\paragraph{Authoring} 
To leave feedback, a coach selects a joint on the avatar or the 2D skeletal diagram, which brings up a floating input panel in front of the avatar, then records a voice note or types a text annotation. The completed annotation is rendered as a small orange marker on the timeline so it can be revisited later by scrubbing to it in playback.

\paragraph{Playback}
To review a previously authored session, the coach selects it from a dropdown populated with saved clips that have been reviewed. The system automatically places a marker along the timeline at each saved annotation's timestamp, giving the coach an at-a-glance map of where feedback was left before watching the clip in detail. Direct joint selection and the 2D skeletal diagram are both disabled in this mode, preventing scrubbing and review from unintentionally creating or moving a note. As the coach scrubs to each marked timestamp, the corresponding annotation becomes active, displaying its text or playing its recorded audio at the associated joint. A delete-annotation control appears only while an annotation is active, letting the coach remove a single outdated note without affecting the rest of the session while a separate delete-review control lets the coach discard an entire saved session at once.

\subsection{Deployment}~\label{sys:deploy}
\sys~was implemented in Unity 6000.0.75f1~\footnote{\url{https://unity.com/}} and deployed to a Magic Leap 2 headset. For our prototype implementation and the evaluations reported in this paper, \sys~shipped with a preloaded set of FBX files covering the reference lunge and error exemplars used across both the \textit{Trainee} and \textit{Reviewer} modules (Section~\ref{sys:rationale}). Beyond this preloaded set, new movement replays can be added to either module by placing additional FBX files, exported through the motion capture pipeline (Section~\ref{sys:mocap}), into the application's persistent storage path, without modifying the application itself.

Rather than embedding a learner's or coach's contributions directly within the movement data itself, \sys~persists them as separate artifacts linked back to the originating clip, an approach to structuring user-generated feedback inspired by DanXeReflect~\cite{danxereflect2026}. In the \textit{Trainee} module, each testing-mode decision a learner makes is stored in a separate CSV file containing timestamp, the learner's decision, time taken to make the decision, and their stated confidence to the clip being judged, allowing this data to be recovered for later analysis of learning outcomes. In the \textit{Reviewer} module, a coach's annotations are similarly persisted separate from the movement itself. Each annotation is anchored to a specific body joint and a normalized timestamp within the clip, carries either a text message or a reference to a locally stored voice recording, depending on the annotation type used and is stored as JSON file. This separation allows a completed review to be reloaded and reattached to its clip at any later time, whether by the coach revisiting their own analysis, or by a student replaying the annotated session asynchronously to see each note resurface at its correct spatial and temporal location in the movement.

\section{System Evaluation}

Before evaluating the \textit{Trainee} and \textit{Reviewer} modules with users, we first validated that our motion capture pipeline could reliably reconstruct lunge movements suitable for training and coaching feedback. This section describes the dataset collected for this purpose and the validation confirming that the resulting 3D reconstructions preserved the kinematic features distinguishing correct lunges from shoulder rotation and leg-leading errors. 

\subsection{Evaluation Dataset}\label{sec:dataset}
To create a dataset for system evaluation, we recorded a fencing expert performing lunges in a well-lit university practice studio. A Google Pixel 8 Pro smartphone was used to record the movements in 1080p at 60 frames per second and  placed 3 meters from the fencer's initial position to ensure full-body capture throughout the movement. The expert executed correct lunges as well as lunges with shoulder rotation and leg-leading errors. These errors were performed with varying degrees of severity, ranging from subtle to pronounced, to prevent ceiling and floor effects in participant performance. The resulting dataset contained approximately 15 minutes of footage and 30 repetitions for each lunge type.

\subsection{Motion Reconstruction Validation}
To quantitatively assess whether reconstruction preserves the kinematic differences that distinguish erroneous from correct lunges, the generated motion sequences were processed using a customized kinematic analysis pipeline designed to extract joint kinematics that serve as discriminating biomechanical markers between correct and erroneous lunges.

As the reconstructed repetition segments varied slightly in spatial orientation and initial position, the joint trajectories were transformed into a canonical coordinate space. The root position of the 3D skeleton was reset to $(0, 0)$ in the transverse $X-Z$ plane and the skeleton was rotated to align its initial facing direction along the positive $Z$-axis. This transformation ensured that the subsequent angular calculations and joint displacements were invariant to the camera viewpoint and initial position of the skeleton. To isolate the active phase of the lunge and to filter out initial preparation and terminal return movements, we temporally restricted the analysis window to the $20\%$ to $95\%$ segment of each repetition. 

From this window, we extracted the peak transverse-plane angle between the shoulder girdle axis and the weapon-side upper arm and the peak leg extension frame normalized by the active analysis window length. The peak transverse-plane shoulder angle was greater during lunges with shoulder rotation error $(M : 171.78^{\circ}; SD : 6.5^{\circ})$ than during technically correct lunges $(M: 160.58^{\circ}; SD : 5.9^{\circ})$. A Mann-Whitney U-test indicated that this difference was significant $(U = 251.0, p < 0.001)$. Similarly, a Mann-Whitney U-test indicated that the normalized peak leg extension frame was significantly higher in the correct lunges $(M = 0.81; SD = 0.28)$ than in the lunges where the leg was leading the arm $(M = 0.28; SD = 0.2; U = 24.0; p < 0.001)$. These results indicate that the reconstruction pipeline preserves the kinematic signal separating each error type from correct execution, rather than establishing the absolute accuracy of the reconstructed joint angles against a ground-truth reference, which this dataset does not include.

\section{Trainee Module User Study}
We conducted a user evaluation to examine whether ~\sys would facilitate short-term improvement in error detection among novice fencers. The study was conducted across two locations, a university laboratory and a fencing practice studio. Both locations were well-lit and had open space of at least $4m \times 4m$ to allow participants to move freely and examine the digital avatar depicting the exemplar movements. 

\subsection{Study Task}

Participants were trained using the system to identify whether a fencing lunge was technically correct or exhibited either a shoulder rotation error or a leg-leading error. These two error types were selected based on the guidance by our second author, a fencing expert, outlined in Section~\ref{sys:rationale}.
All replays used in the study were drawn from the validated dataset described in Section~\ref{sec:dataset}.

\subsection{Study Procedure} 
Participants started by providing informed consent (study protocol approved by \#anonymous IRB) and demographic information. Then they completed the Affinity for Technology Interaction (ATI) survey~\cite{franke2019personal} and a custom pre-study questionnaire that assessed their fencing experience, training frequency, familiarity with AR, and usage of video analysis for improvement. 

Participants were then briefed on the objectives of the study and introduced to the two errors in the lunge using instructional videos. They were also introduced to the AR interface and their assigned visualization modality. Following this, participants completed the standard Magic Leap 2 custom fit calibration, which also served as an interactive tutorial to familiarize the them with the headset and the controller. This was followed by the main study which consisted of three-phases: a pre-test, a training phase, and a post-test. 

The pre-test and post-test employed identical structures to evaluate baseline and post-training error detection performance respectively. In both phases, participants observed 3D replays of lunges and were given 15 seconds to classify each movement as technically correct, exhibiting a shoulder rotation error, or exhibiting a leg-leading error. Participants could pause the playback, rotate the virtual avatar, or walk around it freely. However, pausing the playback did not pause the 15-second decision timer. After making a decision, participants reported confidence in their decision on a 5-point Likert scale (1 - not confident at all, 5 - very confident), which had no time constraint. A 3-second visual countdown preceded each lunge presentation to allow participants to prepare. To reduce the chances of cognitive and physical fatigue, participants were given a 10-second rest period after every set of 5 replays. Both the pre-test and post-test contained 12 clips consisting of an equal distribution across the lunge types (4 correct lunges, 4 shoulder rotation errors, and 4 leg-leading errors), randomized independently for each participant. Pre-test, training, and post-test clips were drawn from disjoint, non-overlapping subsets of the repetitions in our dataset, so no participant saw the same repetition more than once across the study.

During the training phase, participants watched 21 lunge replays augmented with their assigned visualization modality along with an explicit error label. Participants had 45 seconds per movement to analyze the augmented replay and identify the biomechanical markers associated with the lunge type. Similar to the testing phases, participants could rotate the avatar, walk around it, and pause the playback. The training phase also used the same 3-second pre-clip countdown and 10-second rest break after every 5 clips. Following the training phase and prior to the post-test, participants completed a post-training questionnaire and watched a 5-minute publicly available YouTube video featuring humorous pets. After watching the video, participants had to mention which pet they found the funniest. The objective of this distractor task was to engage the working memory of the participants and prevent the repetition of the learned biomechanical features of the movements before the test~\cite{posner1966short, peterson1959short}. The post-training questionnaire included 7-point Likert scales to assess the interpretability and distractibility of the visualizations, along with their effect on perceived learning and associated extraneous cognitive load (ECL). The ECL items were adapted from ~\cite{klepsch_development_2017}.   

\begin{figure}
    \centering
    \includegraphics[width=\linewidth]{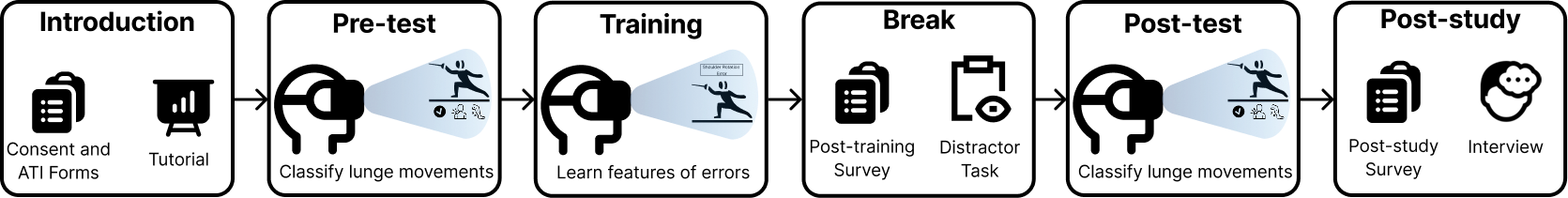}
    \caption{Study procedure: participants provided consent and completed the ATI questionnaire~\cite{franke2019personal}, then were briefed on the study, introduced to the two lunge errors, and set up with the AR interface, their assigned visualization, and headset calibration. Pre-test: participants classified 3D replays as correct or containing a shoulder or leg-leading error. Training: participants viewed replays augmented with their assigned visualization. After the training, they completed the post-training questionnaire and the distractor task. Post-test: identical format to the pre-test. Participants then completed a post-study questionnaire and a semi-structured interview.}
    \label{fig:study_procedure}
\end{figure}

After completing the post-test phase, participants filled out a post-study questionnaire and participated in a semi-structured interview with the researcher. The study procedure is illustrated in Figure~\ref{fig:study_procedure}.

\subsection{Participants}
A total of 18 participants (9 per visualization modality) were recruited for this between-subjects study through mailing lists and word-of-mouth. As the primary objective of this preliminary evaluation was to investigate whether novice fencers could acquire error-detection skills using either visualization modality, we did not conduct a formal apriori power analysis. Participants (8 female, 8 male, 1 non-binary and 1 declined to state) aged between 18 to 30 ($M = 21; SD = 3.6$) years had normal or corrected-to-normal vision and reported no color vision deficiencies. Participants were randomly assigned to a visualization modality. Eleven participants from were from a university fencing club in the United States and were compensated \$20 for their time while rest of the participants were from a fencing club in Bengaluru, India and they were compensated 1000 INR for their time. Each study session lasted approximately 60 minutes. 

Though participants were recruited from fencing clubs, their self-reported fencing experience indicated that they were novices. Six participants had less than six months of fencing experience, six had between 6 and 12 months, four had between 1 and 2 years, and 2 had more than 2 years of experience but were returning to the sport after a break of greater than six months. The baseline performance (Accuracy $M = 37.5\%, SD = 14.6$, close to chance $33\%$) from the pre-test indicated that there was room for performance improvement for all the participants.

\subsection{Quantitative Results}~\label{sec:ed_quant}
To assess the effects of training on participant's error-detection abilities, we examined three categories of dependent variables: (a) objective error-detection performance, captured using accuracy, decision time, no response rate, and confidence, (b) self-reported perceived learning, and (c) extraneous cognitive load (ECL)~\cite{klepsch_development_2017}, both measured via post-training survey.
\subsubsection{Error-detection Performance}

\begin{figure}
    \centering
    \begin{subfigure}[b]{0.495\linewidth}
        \centering
        \includegraphics[width=\textwidth]{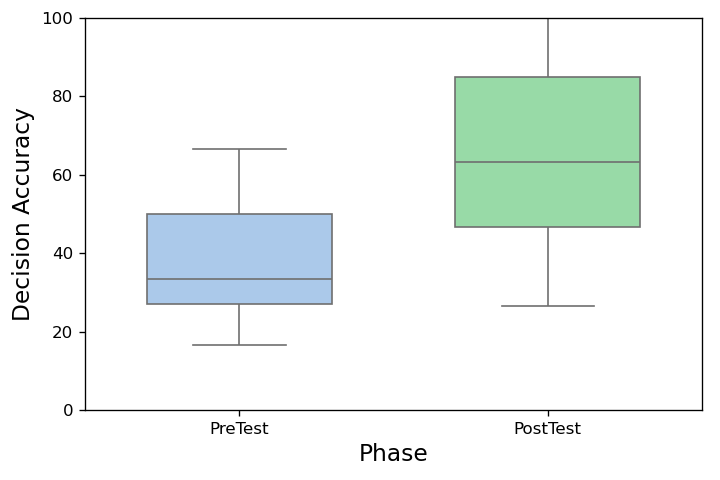}
        \caption{Decision-making accuracy before and after training with ~\sys~across both modalities}
        \label{fig:acc_pre_post}
    \end{subfigure}
    \begin{subfigure}[b]{0.495\linewidth}
        \centering
        \includegraphics[width=\textwidth]{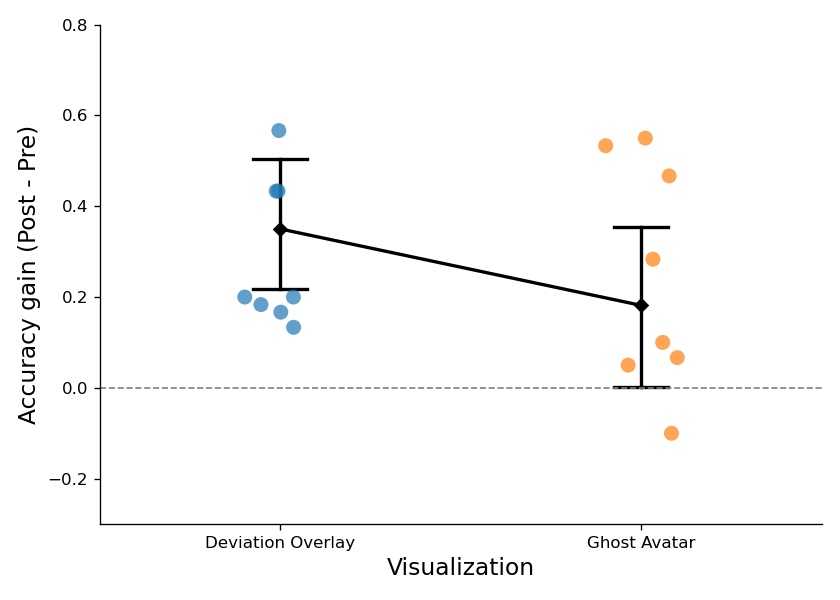}
        \caption{Individual learning gains per participant}
        \label{fig:learning_gain}
    \end{subfigure}
    \caption{(a) Decision-making accuracy from unaided pre-test to post-test for the \textit{Deviation Overlay} and \textit{Ghost Avatar} visualization modalities. Accuracy increased from pre- to post-test across both modalities, indicating that both visualizations supported learning. (b) Individual accuracy gains were slightly higher for participants in the \textit{Deviation Overlay} group than for those in the \textit{Ghost Avatar} group but differences were not significant.}
    \label{fig:perf_metrics} 
\end{figure}

\paragraph{Accuracy}
We computed error detection per participant, per phase (pre-test, post-test), and per visualization modality (\textit{Deviation Overlay}, \textit{Ghost Avatar}). Trials where participants failed to respond within the 15-second response window were considered as an incorrect decision, reflecting a failure to correctly discriminate the movement under time pressure. As a Shapiro-Wilk test indicated that the accuracy data was normally distributed $(W=0.94, p=0.06)$, we conducted a $2 \times 2$ mixed-design ANOVA, with Phase as a within-subjects factor and Visualization Modality as a between-subjects factor (N = 18; n = 9 per modality). ANOVA results indicated a significant effect of both Phase and Visualization Modality. Accuracy improved significantly from pre-test $(M = 37.5, SD = 14.6)$ to post-test  $(M = 64.07, SD = 23.5 ; F(1,16) = 17.48, p < .001, \eta^{2}p = .52)$, indicating that the participants became more accurate after the training (Figure~\ref{fig:acc_pre_post}). A significant main effect of Visualization Modality $(F(1,16) = 4.82, p = .043, \eta^{2}p = .23)$ indicated an overall accuracy difference between \textit{Deviation Overlay} $(M = 57.3, SD=25.1)$ and \textit{Ghost Avatar} $(M = 44.2, SD = 20.5)$, averaged across phases. However, the Phase $\times$ Visualization Modality interaction was not significant $(F(1,16) = 1.76, p = .204, \eta^{2}p = .10)$, indicating that the magnitude of improvement from pre- to post-test did not differ significantly between visualization modalities. 

 Follow-up paired t-tests within each visualization modality showed significant improvement in accuracy from pre to post-test in \textit{Deviation Overlay} $( pre = 39.8 \pm 16.5, post = 74.8 \pm 19.3; t = -4.43, p = .002, d=1.94)$ but not in \textit{Ghost Avatar} $(pre = 35.2 \pm 13.02, post = 53.3 \pm 23.3  ; t = -1.82, p = .11, d=0.96)$. 
Further, a post-hoc comparison of post-test accuracy showed a marginally significant difference between the visualizations $(T = -2.12, p = 0.05, d=1.0)$. However, the direct between-groups comparison of average improvement magnitude was not significant $(\delta_{Ghost Avatar} =18.1 \pm 29.9; \delta_{Deviation Overlay} = 35.0 \pm 23.6; T = 1.32, p = .20)$. The individual learning gains of each participant across the two modalities are shown in Figure~\ref{fig:learning_gain}.

\paragraph{No Response Rate} 
The proportion of trials with no response decreased significantly from pre-test $(M = 0.13, SD = 0.11)$ to post-test $(M = 0.04, SD = 0.05; F(1,16) = 14.56, p = .002, \eta^{2}p = .48)$, with a non-significant trend toward more no response trials in \textit{Ghost Avatar} $(M=0.11, SD = 0.1)$ compared to the \textit{Deviation Overlay} visualization $(M = 0.04, SD=0.04;  F(1,16) = 3.75, p = .071, \eta^{2}p = .19)$. Phase $\times$ Visualization Modality interaction was not significant $(p = .129)$.

\paragraph{Confidence}
Considering only the trials in which a decision was made, the participant's confidence in their decision increased significantly from pre- to post-test $(conf_{pre} = 3.44 \pm 0.73, conf_{post} = 3.85 \pm 0.71 ;F(1,16) = 17.27, p < .001, \eta^{2}p = .52)$. Both the Visualization Modality $(conf_{ghost} = 3.55 \pm 0.82, conf_{deviation} = 3.74 \pm 0.66; F(1,16) = 0.33, p = .571)$ and Phase $\times$ Visualization Modality $(p=0.8)$ did not have a significant effect on the decision confidence.

\paragraph{Decision Time}  
On trials with a decision, time taken to make the decision did not differ significantly between pre- and post-test ($F(1,16)=0.40, p=.537$). Decision time did differ significantly by visualization modality ($F(1,16)=5.25, p=.036, \eta^2_p=.25$), with\textit{Ghost Avatar} participants ($M=8.13\mathrm{s}, SD=1.55\mathrm{s}$) taking longer to make a decision than \textit{Deviation Overlay} participants ($M=7.33\mathrm{s}, SD=1.67\mathrm{s}$) across the two phases.

\subsubsection{Perceived Learning}
In addition to objective error-detection performance, we assessed participant's perceived learning using four 7-point Likert-scale (0–6) items (Table \ref{tab:pl_questions}) within the post-training survey. Perceived learning captures a subjective evaluation of skill acquisition that is distinct from objective accuracy metrics, serving as a key indicator of self-efficacy, continued engagement, motivation, and potential skill transfer~\cite{bandura1997self, rovai2009development}. The overall perceived learning score was computed as the mean response across the four items. Figure~\ref{fig:pl_viz} illustrates these findings. An independent samples t-test revealed no significant difference $(T = 1.07, df=16, p=0.3)$ in perceived learning between the \textit{Ghost Avatar} ($M = 5.02$, $SD = 0.42$) and \textit{Deviation Overlay} ($M = 4.72$, $SD = 0.74$) visualization modalities.

\begin{table}[h!]
    \centering
     \caption{
     Custom questionnaire used to measure perceived learning. Responses were recorded on a 7-point Likert scale.}
    \begin{tabularx}{0.99\linewidth}{c X c c}
        \toprule
        \textbf{\#} & \textbf{Prompt Item} & \textbf{Anchor 1} & \textbf{Anchor 7} \\
        \midrule
         1. & For learning differences between the movement types, this visualization was & Inconvenient & Convenient \\
         2. & The training helped me build a clear understanding of what each error looks like & Strongly Disagree & Strongly Agree \\
         3. & I actively thought about the differences between correct and incorrect technique during training & Strongly Disagree & Strongly Agree \\
         4. & I feel like I developed a mental picture of what each error looks like & Strongly Disagree & Strongly Agree \\
         \bottomrule
    \end{tabularx}
   
    \label{tab:pl_questions}
\end{table}

\subsubsection{Extraneous Cognitive Load}
\begin{figure}
    \centering
    \begin{subfigure}[b]{0.49\linewidth}
        \centering
        \includegraphics[scale=0.5]{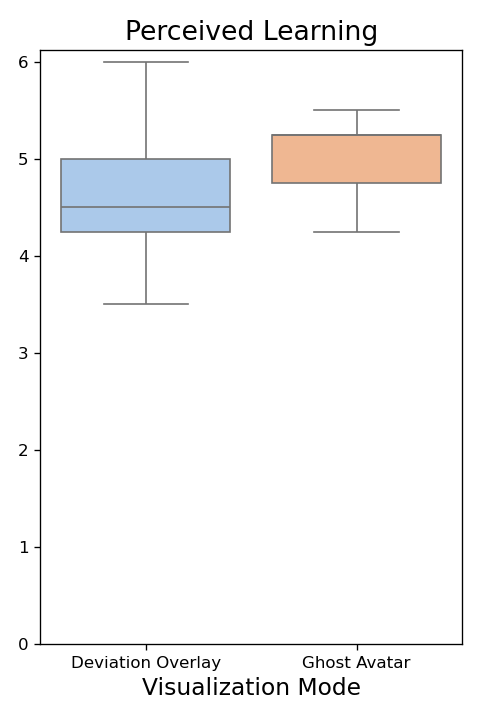}
        \caption{Perceived Learning}
        \label{fig:pl_viz}
    \end{subfigure}
    \begin{subfigure}[b]{0.49\linewidth}
        \centering
        \includegraphics[scale=0.5]{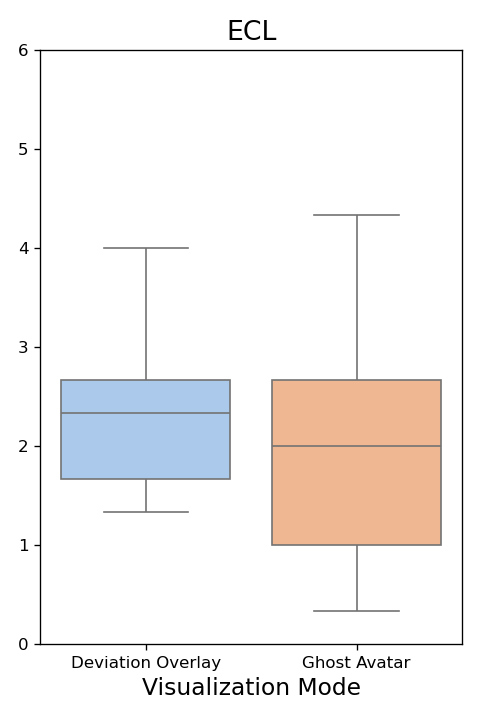}
        \caption{Extraneous Cognitive Load}
        \label{fig:ecl_viz}
    \end{subfigure}
   
        \caption{Perceived Learning (higher is better) and ECL (lower is better) scores for the visualization modalities used in our evaluation. The two modalities produced similar scores on both measures, with high Perceived Learning and low ECL overall, suggesting neither modality imposed excessive cognitive burden relative to the learning it supported.}
    \label{fig:pl_ecl}
\end{figure}

The presentation layout and the instructional design of the learning material or system application induces extraneous cognitive load in learners~\cite{klepsch_development_2017, sweller1994cognitive}. They often have to dedicate their finite cognitive resources to processes such as ignoring irrelevant content or repeatedly searching for information which can impact their learning efficiency~\cite{sweller1994cognitive}. We measured ECL using a three-item, 7-point Likert scale adapted from the validated instrument developed by \citet{klepsch_development_2017} (Table~\ref{tab:ecl_questions}). The overall ECL was computed as the mean of responses to the three questions. The results are shown in Figure~\ref{fig:ecl_viz}. An independent samples t-test revealed no significant difference $(T = -0.59, df=16, p=0.56)$ in ECL induced between the \textit{Ghost Avatar} ($M = 2.03$, $SD = 1.24$) and \textit{Deviation Overlay} ($M = 2.33$, $SD = 0.85$) visualization modalities.

\begin{table}[h!]
    \centering
     \caption{
     Extraneous cognitive load was measured using items adapted from ~\cite{klepsch_development_2017}. These items evaluated the difficulty participants experienced in acquiring the information required to distinguish between different lunge types. Responses were recorded on a 7-point Likert scale ranging from 1 (\textit{Strongly Disagree}) to 7 (\textit{Strongly Agree})}
    \begin{tabularx}{0.99\linewidth}{l X}
        \toprule
        \textbf{\#} & \textbf{Prompt Item} \\
        \midrule
         1. & This visualization is inconvenient for learning what an error looks like \\
         2. & It was exhausting to find the information needed to spot an error in this visualization \\
         3. & This feedback made it difficult to recognize information and link it to the error types \\
         \bottomrule
    \end{tabularx}
   
    \label{tab:ecl_questions}
\end{table}

\subsubsection{Interpretability and Distraction} 
We compared the interpretability and distraction ratings between \textit{Ghost Avatar} and \textit{Deviation Overlay} using independent-samples tests.

\paragraph{Interpretability}
A Shaprio-Wilk test indicated that interpretability ratings were approximately normally distributed ($p = 0.08$), so we conducted an independent-samples t-test. The \textit{Ghost Avatar} was rated as similarly interpretable ($M = 4.22, SD = 1.48$) to \textit{Deviation Overlay} $(M = 3.78, SD=0.97)$ with no significant difference between the visualization modalities $(t= 0.75, p=0.46, d=0.35)$.

\paragraph{Distraction}
A Shapiro-Wilk test indicated that distraction ratings deviated from normality, so we conducted a Mann-Whitney U test. \textit{Ghost Avatar} $(M = 4.33, SD=1.8)$ and \textit{Deviation Overlay} $(M = 4.44, SD=1.01)$ did not differ significantly in perceived distraction ratings $(U  = 43.5, p=0.82, RBC=0.074)$.


\subsection{Qualitative Results}\label{sec:ed_qual}
After the post-test, participants provided written responses to the question asking what type error identification strategy the system prompted and if it was different from strategy they used currently.  We also conducted semi-structured interviews (Table~\ref{tab:ssi}) with all the participants after they filled out the post-study questionnaire. Interviews were audio-recorded and transcribed by the researcher and responses from non-English-speaking participants were translated by the researcher at the time of transcription. The participants were prompted to evaluate the system, elaborate on their specific learning strategies, and contrast the platform with the technological tools they currently leverage in practice. The responses were analyzed using inductive thematic analysis to systematically code and organize the data into themes~\cite{braun2006using}. 
\begin{table}[t]
\centering
\caption{Questions used for the semi-structured interview conducted after the study.}
\label{tab:ssi}
\begin{tabularx}{\linewidth}{>{\bfseries}l X}
\toprule
\# & \textbf{Question} \\
\midrule

1 &
What was it like to use the training system? \\

2  &
What was your learning strategy during the training phase? \\

3 &
Were there particular aspects of the virtual fencer's movements that helped you make your decision? Did this change between the pre-test and post-test? \\

4 &
In what ways did the 3D motion replay differ from watching a conventional 2D video or observing others during a training session? \\

5 &
How do you think this system would fit into your current training or review practices (e.g., video analysis, mirrors, or coaching sessions)? \\

\bottomrule
\end{tabularx}
\end{table}

\subsubsection*{\textbf{Theme 1:  Spatial and Temporal Control}}~\label{sec:t1}

Participants across both visualization modalities identified control over the spatial perspective and temporal progression of the movement replay as the main advantage of ~\sys over 2D video analysis and live observation. 

\paragraph{Multi-angle viewing} 
~\sys's rotation control let participants match the replay viewpoint to the mechanic they were checking. They developed a consistent strategy of rotating to a side view to check leg timing or foot placement, then to a front view to check shoulder alignment. This strategy emerged during training and persisted into the post-test. P7 (\textit{Deviation Overlay}) mentioned that they ``like the side angle for the leg leading, and the front angle for the shoulder rotation''. Some participants explored different viewing angles and ultimately settled on one. P9 (\textit{Deviation Overlay}) mentioned that they were a ``big fan of the like 45 degree'' as that how they normally view fencers in their club and P1 (\textit{Deviation Overlay}) mentioned that they preferred the side view ``just like the judge". Some participants (P3, P9) noted a preference for rotating the avatar in place using the UI rather than walking around it.

\paragraph{Pausing and looping} 
Fencing actions happen too quickly to assess in real time, particularly for beginners. Looping the replay and pausing it let participants carefully analyze specific moments rather than relying on a single pass. P4 (\textit{Ghost Avatar}) explained that detecting errors from live action is demanding because it is ``so fast sometimes,'' requiring them to ``really watch the move and think about it'', whereas pausing the replay and changing the angle made the analysis simpler. P8 (\textit{Ghost Avatar}) described a deliberate pause-and-inspect strategy, pausing wherever they thought the mistake was occurring, and if the movement still looked correct at that point, pausing elsewhere and trying again.

\paragraph{Consistency across repetitions}
Many participants (P1, P2, P5, P9) noted that coaches and peers cannot reproduce the same error identically across repetitions, so live training does not give learners a standardized example to learn from. The avatar's playback was identical on each repetition, making it easier to isolate the biomechanical markers of the error being studied. P9 (\textit{Deviation Overlay}) noted that in real life, a repeated mistake still looks ``a different amount of bad'' each time it occurs, making it harder to learn to identify.

\subsubsection*{\textbf{Theme 2 : Visualizations and Learning Strategy}}~\label{sec:t2}
\sys's visualizations effectively guided participant's attention during training, but participants used the augmentations in the two modalities differently.

\paragraph{Ghost Avatar: comparison strategy}
\textit{Ghost Avatar} participants described using a comparison strategy, systematically checking the ghost avatar against the main avatar to locate positional discrepancies. P14 (\textit{Ghost Avatar}) mentioned that they went ``from the upper body all the way to the feet," and P16 (\textit{Ghost Avatar}) reported moving ``joint by joint'' between the two avatars. The side-by-side layout of the \textit{Ghost Avatar} visualization supported a sequential, body-part-by-body-part inspection. P2 (\textit{Ghost Avatar}) indicated that they liked seeing the proper technique shown next to the error, noting that having a ``comparison like right there was pretty helpful for learning what's the bad action being done here.''

\paragraph{Deviation Overlay: label-guided reading strategy}
Deviation Overlay participants more often described a reading strategy, with the error labels functioning as an initial anchor for attention with P17 (\textit{Deviation Overlay}) reported first looking at the label and then using the error indicators to focus on specific parts of the fencer, rather than scanning the body generally. Similarly, P5 (\textit{Deviation Overlay}) reported anchoring their attention on specific joint markers,  ``I focused more on the elbow, wrist, and shoulder visualization, the yellow dots.''  P1 (\textit{Deviation Overlay}) used what they called the ``ribbon paths'' -- the joint displacement line segments traced across successive frames of the replay -- to connect arm-leg timing to a broader issue with core balance: ``I feel like I can understand the unbalance, because... the leg [is] reaching out first, and then the arms are reaching out further, and then that's the unbalance of the core.''


\paragraph{Limitations of the Visualizations}
While many participants (P2, P5, P7, P8) liked the visualizations and found it helpful, few participants (P4, P17) reported some cognitive friction. P4 (\textit{Ghost Avatar}) found it confusing when the ghost avatar and the main avatar would "glitch into each other or like superimpose." P17 (\textit{Deviation Overlay}) similarly found the ``error-indicator spheres confusing'' and less intuitive than a directional cue would have been.

\subsubsection*{\textbf{Theme 3: From Guessing to Structured Analysis}}~\label{sec:t3}
Participants across both visualization modalities described moving away from broad, instinct-driven scanning toward structured inspection of specific joints and kinetic relationships.

\paragraph{From holistic scanning to inter-joint timing}
Before training, many participants (P1, P5, P7, P12, P18) reported looking at the whole body or restricting their attention to the weapon arm, relying on general experience rather than a specific strategy. Post-training, this shifted toward monitoring not just the location of individual joints but the relative timing between body segments such as checking whether arm extension occurred in the correct cadence with the movement of the lead leg. P1 (\textit{Deviation Overlay}) described this shift, mentioning on the pre-test they ``just tried to base [answers] on my experience,'' whereas by the post-test they had started noticing ``the hands reaching out first, or the leg reaching out first'' and connecting that timing to the resulting imbalance. P7 (\textit{Deviation Overlay}) reported a similar narrowing, moving from ``much more broad'' scanning while still figuring out what to look for, to focusing on specific joints and working ``much quicker'' by the post-test.
This narrowing was accompanied by a redistribution of attention. Several participants (P4, P10, P12, P13, P14) noted they began observing joints they had previously overlooked. P12 (\textit{Ghost Avatar}) described this shift concretely, noting their observation moved ``from full body to specific body parts such as thigh, forearm.'' P4 (\textit{Ghost Avatar}) reported focusing ``mainly at the leg'' during the pre-test but attending to additional joints by the post-test, and P13 (\textit{Deviation Overlay}) similarly went from ``ignoring leg movements'' to noticing them during the post-test.

Finally, two participants (P6, P8) mentioned that the training phase helped reinforce their understanding of the errors. 

\paragraph{Kinetic Chain and Center-of-Gravity Awareness}
Beyond recognizing the incorrect movement of individual joints, \sys~also helped participants (P1, P6, P9, P15) connect localized errors to their effects down the kinetic chain. P6 (\textit{Ghost Avatar}) mentioned they now understood how technical errors led to disadvantageous shifts in ``center-of-gravity'', P9 (\textit{Deviation Overlay}) similarly noted how shoulder rotation resulted in the ``wrist being out of place''. 

\subsubsection*{\textbf{Theme 4: \sys~ as a Complement to Existing Tools }}~\label{sec:t4}
Participants consistently positioned \sys~as a complement to human coaching. They viewed it as a high-fidelity diagnostic tool for autonomous practice and reflection between coaching sessions. 

\paragraph{Cognitive Priming and Reflection}
The most common proposed use cases for \sys~were independent pre-lesson self-review (P7, P15), self-directed error identification between coaching sessions (P13, P16), and post-practice reflection (P5). P7 (\textit{Deviation Overlay}) mentioned they would use the system to ``prime my brain before an in-person lesson'' and build a mental model of errors made by them before discussing with a coach or expert. P16 (\textit{Ghost Avatar}) mentioned they would use the system to train themselves to spot errors in their own movements and consult with a coach after.   

\paragraph{Suitability across experience levels}
Most participants felt \sys~would be most beneficial to beginner and intermediate fencers learning to identify and classify movement errors especially if they could use their own movements with the system (P1, P3, P15). P5 (\textit{Deviation Overlay}) specifically noted its value for those ``still in the starting phase'' of skill acquisition. P16 (\textit{Ghost Avatar}) though cautioned that it might not be that ``useful for beginners (under three months of experience)'', while P4 (\textit{Ghost Avatar}) speculated that ``professionals can study the movements their opponents make in more detail''.

\section{Reviewer Module Study}

To conduct a preliminary evaluation of the \textit{Reviewer} module's interface and workflow, we conducted an asynchronous online study using video prototypes. Using video prototypes allowed us to recruit experienced fencing coaches and experts regardless of location or headset access. The video was recorded using the built-in video capture feature of the Magic Leap 2, with a researcher demonstrating the module's interface elements, examples of text and audio annotations, and playback capabilities. We recruited four fencing experts (2 female, 2 male) through word-of-mouth. We considered fencers with at least 5 years of competitive experience competing at national-level tournaments as experts. Participant details are shown in Table~\ref{tab:exp_part}.

\begin{table}[h!]
    \centering
     \caption{
     Fencing experts that participated in the \textit{Reviewer} module study}
    \begin{tabularx}{0.99\linewidth}{l l l l X}
        \toprule
        \textbf{Participant ID} & Age & Years Fencing & Years Coaching & Location \\ 
        \midrule
         1 & 28 & 10 & 3 & United States of America \\
         2 & 29 & 10 & 2 & United States of America \\
         3 & 48 & 31 & 20 & United States of America \\
         4 & 21 & 8 & 2 & India \\
         \bottomrule
    \end{tabularx}
   
    \label{tab:exp_part}
\end{table}

\subsection{Task} 
Participants watched an approximately 4-minute long video demonstrating the \textit{Reviewer} module's interface, annotation workflow, and playback features. They then rated the usefulness of the system capabilities demonstrated in the video (e.g., avatar rotation, slow-motion replay, motion play/pause), the clarity of the annotation interface (cluttered vs. clear, obstructive vs. unobstructive), and the efficiency of the annotation-creation and playback workflow using Likert-scale items. Participants also indicated whether they viewed the system as a complement or a replacement for their existing coaching tools. They answered open-ended questions about: (1) what interface elements felt confusing, unnecessary, or missing, (2) how localized 3D feedback compared to their current remote-instruction or feedback tools, whether the system would help them notice errors they might miss in video and their general impressions, including favorite and missing features.

\subsection{Results}

Given the small sample size (n=4), we did not conduct a formal thematic analysis. Instead one researcher reviewed the participants' Likert ratings and responses to open-ended questions and coded them to identify points raised by two or more participants. Three of four coaches responded positively to the avatar rotation and pause features, and two specifically highlighted the slow-motion replay as valuable for closely inspecting the lunge. Two coaches liked being able to point directly at joints to anchor feedback.

All four coaches indicated the system would be a good complement, rather than a replacement, to their existing tools. When asked how localized 3D feedback that can be provided using ~\sys compared to the remote-instruction tools they currently use, two coaches elaborated that this capability would let them give more comprehensive feedback as they would be able to spot errors that might not be visible in standard video, and would let them give students feedback that is easier to understand.

Two concerns recurred across responses. First, two coaches felt that typing text annotations would be slow, consistent with known limitations of text entry in XR~\cite{textXR}, though one coach specifically noted they liked seeing the text annotation appear during playback. Second, two coaches felt the animation appeared somewhat choppy and expressed a preference for higher frame-rate replays.This may reflect a limitation of our video prototype rather than the system itself. Playback in the \sys~interface is driven by Unity's animation system and    renders smoothly at the headset's native frame rate, but the video shown to participants was captured via the Magic Leap 2's built-in recorder, which may not fully preserve this frame rate, potentially contributing to the perceived choppiness.


\section{Discussion and Design Implications}
Our results show that \sys~can help learners develop transferable error-detection skill and enable coaches to provide spatially-grounded, easily understandable feedback. The visualization used shaped the strategy learners adopted, and both learners and coaches saw clear roles for such a system alongside, rather than in place of, existing coaching and video-based practice. We interpret these results and derive design guidelines for future XR movement replay and motor skill training systems.

\subsection{Is error detection trainable using XR systems? Did participants learn?}

The improvement in unaided classification accuracy from pre- to post-test (Section~\ref{sec:ed_quant}) suggests that error detection may be trainable through structured observation in AR, without requiring learners to physically repeat the movement themselves. This is consistent with prior evidence that watching another person perform a skill can build error-detection ability as effectively as physical practice~\cite{blandin2000}, and extends that finding to fencing, where the movement is too fast to observe informally and the errors are not visually obvious to a novice.

The qualitative shift participants described, from broad scanning toward targeted inspection of specific joints and their relative timing (Section~\ref{sec:ed_qual}), maps onto the transition motor learning theory locates between the cognitive stage, in which learners attend to a movement's overall pattern, and the associative stage, in which detecting and localizing deviation becomes central~\cite{fitts1967human}. Framed this way, the training phase functioned less as instruction in the specific errors we tested than as a nudge along this general progression, which is consistent with participants who described training as reinforcing an existing intuition.

Beyond the visualization differences, the system's core capabilities of freely rotating the avatar and pausing, looping and replaying the movement identically each time, may have contributed to learning (Theme 1; Section ~\ref{sec:t1}). These capabilities directly addressed limitations of live demonstrations and video review outlined in Section~\ref{sys:rationale}. While \textit{Deviation Overlay} showed a significant pre-to-post improvement in accuracy and \textit{Ghost Avatar} did not, the Phase $\times$ Visualization Modality interaction was not significant, indicating the magnitude of improvement did not differ significantly between the two modalities. We therefore do not interpret this as evidence that one visualization produced a meaningfully larger learning benefit than the other. Instead, both visualization modes appear to have contributed to comparable amounts to learning but directed attention differently, with the system's core capabilities enabling that attention to be used productively.

Some participants went further than simple classification, connecting a detected error to a downstream consequence, such as poor balance or unintended shift in center-of-gravity, without being asked to do so. This complements the accuracy improvement as indicator of learning, suggesting participants were building a working model of the movement rather than just isolating error-defining biomechanical features. We treat this as a promising signal, especially since our system never showed participants their own movement. Whether this reasoning would transfer to correcting a learner's own lunge remains an open question for future work. 

Our results provide preliminary evidence that XR systems can train error-detection, a perceptual skill known to support motor skill acquisition~\cite{seidler2013}. Hence, we recommend that future XR systems for motor skill training should consider incorporating a dedicated error-detection training module, alongside physical practice, rather than focusing on improving movement accuracy alone.

\subsection{Visualizations for Error-Detection Training}

The two visualizations led to different learning strategies. \textit{Ghost Avatar} prompted active comparison between the reference and exemplar avatars, while Deviation Overlay gave participants a faster, label-anchored read, which matches the design of the visualizations. The pre-to-post gain reached significance within Deviation Overlay but not within \textit{Ghost Avatar}. The Phase $\times$ Modality interaction, however, was not significant, so we cannot interpret this as evidence that one modality produced a larger learning benefit than the other. Both visualizations appear to have supported comparable amounts of learning, while directing attention in different ways.

Self-reports do not align with performance data. Perceived learning, interpretability, distraction, and cognitive load did not differ between modalities, and neither did the magnitude of the accuracy gain itself. Participants' sense of what helped them did not track what we measured in either direction, a pattern consistent with the motor learning literature on judgments of learning, where self-reported confidence in one's
own learning can shift independently of actual performance gains~\cite{kantak2012learning, soderstrom2015learning}. This argues for weighing behavioral data over stated preference when
evaluating a system like ours.

The shortcomings in visualization design that participants mentioned suggest a fix for future XR error-training systems. \textit{Deviation Overlay} could keep its explicit indicator, but add a directional cue, an arrow or gradient showing which way a joint should move, to resolve the ambiguity participants reported without losing the lower attentional burden that that appears to have made its detection strategy faster to apply.

Since the two visualizations appear to support different learning strategies, designers should consider a progressive visualization scheme that introduces direct cues first and indirect, comparison-based practice later may help learners build an initial schema before being asked to apply it independently. This mirrors the use of faded worked examples, shown to be effective in skill learning~\cite{renkl2004fading}.

\subsection{Complementary Use and Adoption Barriers}
Participants consistently framed \sys~as a complement to coaching, not a replacement for it. The use cases they proposed, priming before a lesson, practicing between sessions, and reflecting afterward, cluster around moments when a coach is not available or present. This matches the access problem that motivated our design: coach attention is scarce and concentrated in class time, and our system targets the time outside it. That participants reached this framing on their own, without being prompted with our motivating scenario, indicates that the problem we identified is one learners face in their own training.

This complementary framing was echoed independently by coaches and experts in our video-walkthrough study. Two of four experts felt the system would be a valuable addition to their existing video-analysis toolkit, for two reasons: it would let them catch errors that are difficult or impossible to identify from standard video, and it would let them communicate feedback to students in a form that was easier to understand than verbal description or a single video clip. This suggests the system's value is not confined to independent student practice. Coaches saw a role for it within their own instructional process, using the replay as a shared, precise reference point during or around a lesson.

Two factors, however, currently limit \sys~from displacing existing video-based analysis entirely. First, unfamiliarity with AR/VR headsets was raised as a practical barrier to adoption (P17); for fencers accustomed to reviewing footage on a phone or laptop, the overhead of putting on and operating unfamiliar hardware is a real cost the system needs to justify, independent of its training value. Second, several participants wanted to import their own recorded movement into the system rather than relying only on the built-in exemplar and error library (P1, P3, P15). This request would extend the system from exemplar-only replay to training on a learner's own body, directly connecting to the self-correction that is critical for effective motor skill acquisition~\cite{duke2009s}. The current evaluation demonstrates learning to classify a library of pre-defined errors, but not yet the ability to detect and correct errors in one's own technique. Supporting personal motion capture is a natural and participant-requested next step toward closing that gap.

Participants recommended the system for beginner and intermediate fencers, which matches our framing that error detection matters most while a learner's perceptual template is still forming. One participant (P4) suggested advanced fencers might instead use the system to study an opponent's technique, a distinct use case worth exploring in future work rather than assuming the same design serves both populations equally well.

\section{Limitations and Future Work}

Our evaluations show that \sys~enables learners to develop error-detection skill and would be valuable for experts to provide joint-anchored, interpretable feedback to students. However, several limitations qualify our work and motivate future work.

First, our \textit{Trainee} module study aimed to test whether the visualization modalities used in the system would help trainees learn. While the results indicate that novices did learn to spot errors, we cannot determine which specific aspects of the visualizations, or of the system more broadly, were responsible for this learning. Further, our sample size of 18 novice fencers is modest, and several between-visualization comparisons (e.g., post-test accuracy, no-response rates) reached only marginal significance. A future study with an expanded participant pool and a no-visualization control is needed to assess the true contribution of the visualizations to error-detection training.

Second, training in our study consisted of a single session, and the \textit{Trainee} module used a fixed library of pre-recorded exemplar errors rather than the participant's own movement. Several participants explicitly requested the ability to import their own recorded lunges for comparison against expert technique. This is a natural next step, and would let us test whether the error-detection skill we observed, and the kinetic reasoning some participants demonstrated when connecting an error to a downstream consequence like core imbalance, transfers to identifying and correcting errors in a learner's own movement rather than only in a generic exemplar. Longer-term or multi-session studies are also needed to determine whether learning gains persist, and whether comparison-based learning strategies like those elicited by the \textit{Ghost Avatar} visualization become more effective with additional practice time.

Third, our expert evaluation of the \textit{Reviewer} module was conducted through a video walkthrough rather than hands-on use in the AR headset itself. While this let us gather preliminary feedback on the annotation workflow, coaches did not experience the system's spatial and interactive qualities directly, and some of their concerns, such as animation smoothness and preferred frame rate, may be less salient in first-person AR use than in the video walkthrough we conducted. Future work should evaluate the \textit{Reviewer} module through in-person use by coaches and experts working with students over time, ideally in situ during or around actual coaching sessions.

Finally, future work should explore whether \sys's approach extends to other fast, technique-driven sports with similar coaching-access and video-limitation problems, and whether advanced practitioners can use the same 3D replay approach for other purposes, such as opponent analysis, as one participant suggested.

\section{Conclusion}

We presented \sys, an AR system that reconstructs 3D movement replays from monocular smartphone video to support both error-detection training for novices and spatially-grounded feedback authoring for coaches. Across an 18-participant trainee study and a four-expert review study, we found that a single training session significantly improved novices' unaided error-detection accuracy. It also brought about a qualitative shift from broad visual scanning toward targeted inspection of specific joints and their relative timing, and in some cases toward reasoning about the downstream biomechanical consequences of an error. The two visualization modalities we tested, \textit{Ghost Avatar} and \textit{Deviation Overlay}, elicited different observation strategies, comparison versus direct reading, without producing a reliably different amount of learning, suggesting that a system's core capabilities for viewpoint control and repeatable, pauseable playback may matter as much as the specific visualization layered on top of them. Experts consistently saw the \textit{Reviewer} module  as a valuable complement to their existing coaching and video-analysis tools, particularly for feedback that is difficult to communicate through standard video. Together, these findings suggest that grounding movement-based training and feedback directly in a controllable, repeatable reconstruction of the movement itself is a promising direction for AR systems supporting perceptual-skill training. We hope our findings support future work in this space.

\bibliographystyle{ACM-Reference-Format}
\bibliography{eics}

\end{document}